\documentclass[a4paper,twoside,10pt]{letter}
\usepackage{graphicx,saj,multicol,subeqnarray,epsfig,rotate,lscape}

\def\papertype{}

\def\udc{...}
\begin{document}
\baselineskip=3.1truemm
\columnsep=.5truecm
\newenvironment{lefteqnarray}{\arraycolsep=0pt\begin{eqnarray}}
{\end{eqnarray}\protect\aftergroup\ignorespaces}
\newenvironment{lefteqnarray*}{\arraycolsep=0pt\begin{eqnarray*}}
{\end{eqnarray*}\protect\aftergroup\ignorespaces}
\newenvironment{leftsubeqnarray}{\arraycolsep=0pt\begin{subeqnarray}}
{\end{subeqnarray}\protect\aftergroup\ignorespaces}
\newcommand{\D}{$^\circ$}
\newcommand{\HI}{H\,{\sc i}}
\newcommand{\NaI}{Na\,{\sc i}}
\newcommand{\HeI}{He\,{\sc i}}
\newcommand{\HeII}{He\,{\sc ii}}
\newcommand{\FeVII}{[Fe\,{\sc vii}]}
\newcommand{\FeVI}{[Fe\,{\sc vi}]}
\newcommand{\FeV}{[Fe\,{\sc v}]}
\newcommand{\FeI}{Fe\,{\sc i}}
\newcommand{\FeII}{[Fe\,{\sc ii}]}
\newcommand{\FeIII}{[Fe\,{\sc iii}]}
\newcommand{\HII}{H\,{\sc ii}}
\newcommand{\CaII}{Ca\,{\sc ii}}
\newcommand{\hii}{h\,{\sc ii}}
\newcommand{\SII}{[S\,{\sc ii}]}
\newcommand{\SVI}{[S\,{\sc vi}]}
\newcommand{\SIII}{[S\,{\sc iii}]}

\newcommand{\OI}{[O\,{\sc i}]}
\newcommand{\OII}{[O\,{\sc ii}]}
\newcommand{\OIII}{[O\,{\sc iii}]}
\newcommand{\NII}{[N\,{\sc ii}]}
\newcommand{\NeIII}{[Ne\,{\sc iii}]}
\newcommand{\NeIV}{[Ne\,{\sc iv}]}
\newcommand{\ArIII}{[Ar\,{\sc iii}]}

\newcommand{\SQR}{$^{2}$}

\newcommand{\FRII}{FR\,{II}}
\newcommand{\FRI}{FR\,{I}}
\def\p0{\phantom{0}}
\newcommand{\tbsp}{\rule{0pt}{10pt}}
\def\arcmin{\hbox{$^{\prime}$}}
\def\arcsec{\hbox{$^{\prime\prime}$}}
\def\newblock{\hskip .11em plus .33em minus .07em}



\markboth{\eightrm RADIO DETECTION OF 18 RASS \mbox{BL~LAC} OBJECTS
}
{\eightrm M.W.B. ANDERSON and M.D. FILIPOVI\'C }

{\ }

\publ

\type

{\ }


\title{RADIO DETECTION OF 18 RASS \mbox{BL~LAC} OBJECTS}


\authors{M. W. B. Anderson$^{1,2}$ and M. D. Filipovi\'c$^{2}$ }

\vskip3mm

\address{$^1$Sydney Observatory, PO Box K346, Haymarket, Sydney, NSW 1238, Australia}

\address{$^2$University of Western Sydney, Locked Bag 1797\break Penrith South, DC, NSW 1797, Australia}


\dates{January 16, 2009}{TBA, 2009}


\summary{
We present the radio detection of 18 \mbox{BL~Lac} objects from our survey of over 575 deg\SQR~of sky. These 18 objects are located within $20\arcsec$ of the \mbox{X-ray} position, of which 11 have a measured red-shift. All candidates are radio emitters above $\sim$1~mJy and fall within the range of existing samples on the two colour, \mbox{$\alpha_{\rm ro}$ vs $\alpha _{\rm ox}$,} diagram with a transitional population of three (3) evident. Two unusual sources have been identified, a candidate radio quiet BL~Lac, \mbox{RX~J0140.9--4130}, and an extreme HBL, RX~J0109.9--4020, with \mbox{$Log\left( \nu _{\rm peak} \right)\approx19.2$}. The \mbox{BL~Lac} \mbox{Log(N)--Log(S)} relation is consistent with other samples and indicates the \textit{ROSAT} All Sky Survey (RASS) could contain ($2000{\pm}400$) \mbox{BL~Lac} objects. 
  } 



\begin{multicols}{2}
{


\section{1. INTRODUCTION}
 \label{intro}

\mbox{BL~Lac} objects are a rare class of active galactic nuclei
(AGN, see reviews by Kollgaard~(1994), comprising only
$\sim$1.4\% of the total AGN population (\mbox{BL~Lac}
known $\sim$1122, all AGN $\sim$95971, Veron-Cetty \& Veron~2006). 
Standard AGN unification models explain a \mbox{BL~Lac} as a 
near line of sight, non-thermal (i.e. broadband synchrotron 
emission combined with inverse Compton scattering at higher 
frequencies), relativistic jet (i.e. Doppler boosting) produced 
by a \FRI\ radio galaxy nucleus. Consequently, the optical
spectrum dominated by continuum emission from the jet is characterised by
weak emission or absorption lines. This makes the detection of
\mbox{BL~Lac}s difficult at optical wavelengths using conventional 
AGN selection techniques. Most of the existing \mbox{BL~Lac} 
samples have been initially selected at radio or \mbox{X-ray} 
wavelengths using large area sky surveys, followed by time consuming 
optical spectroscopy to identify the \mbox{BL~Lac}'s in the sample. This
led to the conclusion that \mbox{BL~Lac} characteristics are bimodal, divided 
into radio selected \mbox{BL~Lac}s (RSBL; i.e. the 1~Jy sample of
Stickel~et~al.~1991) and X-ray selected \mbox{BL~Lac}'s (XSBL, i.e. the
Einstein Extended Medium Sensitivity Survey (EMSS) sample of
Rector~et~al.~2001).

A model proposed by Giommi \& Padovani (1994) unifies \mbox{BL~Lac}'s as 
a single population, characterised by a spectral energy 
distribution (SED) with peak or cutoff of the synchrotron 
emission in the IR/optical (RSBLs) or in the UV/X-ray (XSBLs). 
To avoid confusion \mbox{BL~Lac}  studies use the classification 
suggested by Padovani \& Giommi (1995) of low energy peak \mbox{BL~Lacs} 
(LBLs, mostly RSBL) or high energy peaked \mbox{BL~Lacs}  
(HBLs, mostly XSBL). Both classes exhibit similar
extreme ``Blazar'' activity (see Urry \& Padovani~1995) --- rapid,
irregular variability, high optical polarization, radio core
dominated morphology (Laurent-Muehleisen~et~al.~1993), flat radio
spectra, super-luminal motion and emission from the radio to gamma ray
bands (Sambruna~et~al.~1996). Given the EMSS and 1~Jy samples represent 
extremes of the \mbox{BL~Lac} population Giommi \& Padovani~(1994) predicted 
the existence of \mbox{BL~Lac}s with properties intermediate to those 
of LBLs and HBLs --- termed intermediate \mbox{BL~Lac}s (IBLs). 

Currently our understanding of \mbox{BL~Lac} objects is far from
complete, due mainly to their rarity limiting the size of
statistically complete samples. One area that is poorly understood
is the luminosity evolutionary behavior of \mbox{BL~Lac}s. While other classes
of AGN show a strong trend of positive evolution, \mbox{BL~Lac}s do
not. LBLs and IBLs exhibit no or a slight positive evolution
(Stickel~et~al.~1991}; Rector \& Stocke~2001; Laurent-Muehleisen~et~al.~1999) 
while HBLs exhibit a negative evolution (Rector~et~al.~2001), meaning they 
are less luminous/frequent in the past than now. Although the properties of
HBLs and LBLs are different, they are not sufficient to
determine if the properties reflect two different populations of
AGN's or the extremes of a single continuous AGN population. The
role of IBLs remain uncertain (Rector~et~al.~2003) as their
observational properties are not well understood due to the small 
number known (Laurent-Muehleisen~et~al.~1999). 

In the last 18 years (since the launch of {\it ROSAT}), the number of 
\mbox{BL~Lac}s has doubled every six to seven years (Fig.~1) as a direct
result of efficient selection techniques (Schachter~et~al.~1993; 
Wolter~et~al.~1997). Stocke~et~al.~(1989) noted that since
\mbox{BL~Lac}s are radio and \mbox{X-ray} loud, their broadband
spectral properties could be used to select out candidate
\mbox{BL~Lac} objects. In a purely \mbox{X-ray} survey, efficiencies
of \texttt{<}5\% are reached (Stocke~et~al.~1991), increasing to
$\sim$10\% at \mbox{F$_{\rm X}$\texttt{>}10$^{-12}$ ergs
cm$^{-2}$ s$^{-1}$} (Perlman~et~al.~1996). In a purely radio survey
efficiencies of $\sim$7\% are reached
(Stickel~et~al.~1991), increasing to about 12\% if flat spectrum
sources are pre-selected. In a combined radio and \mbox{X-ray}
survey efficiencies of 10\% to 20\% or higher can be reached
(Schachter~et~al.~1993; Wolter~et~al.~1997). Recently,
positional correlation of radio and \mbox{X-ray} catalogues has been
undertaken: Deep \mbox{X-ray} Radio Blazar Survey
(Perlman~et~al.~1998), the REX Survey (Maccacaro~et~al.~1998) and the
\textit{ROSAT} All Sky Survey -- Green bank Survey (RGB; 
Laurent-Muehleisen~et~al.~1999). A large sample of 501 radio-selected 
\mbox{BL~Lac} candidates has been compiled by Plotkin~et~al.~(2008) using 
Faint Images of the Radio Sky at Twenty-Centimetres (FIRST) radio 
survey in combination with the Sloan Digital Sky Survey (SDSS) optical spectroscopy. 

\vskip.5cm
\centerline{\includegraphics[width=75mm]{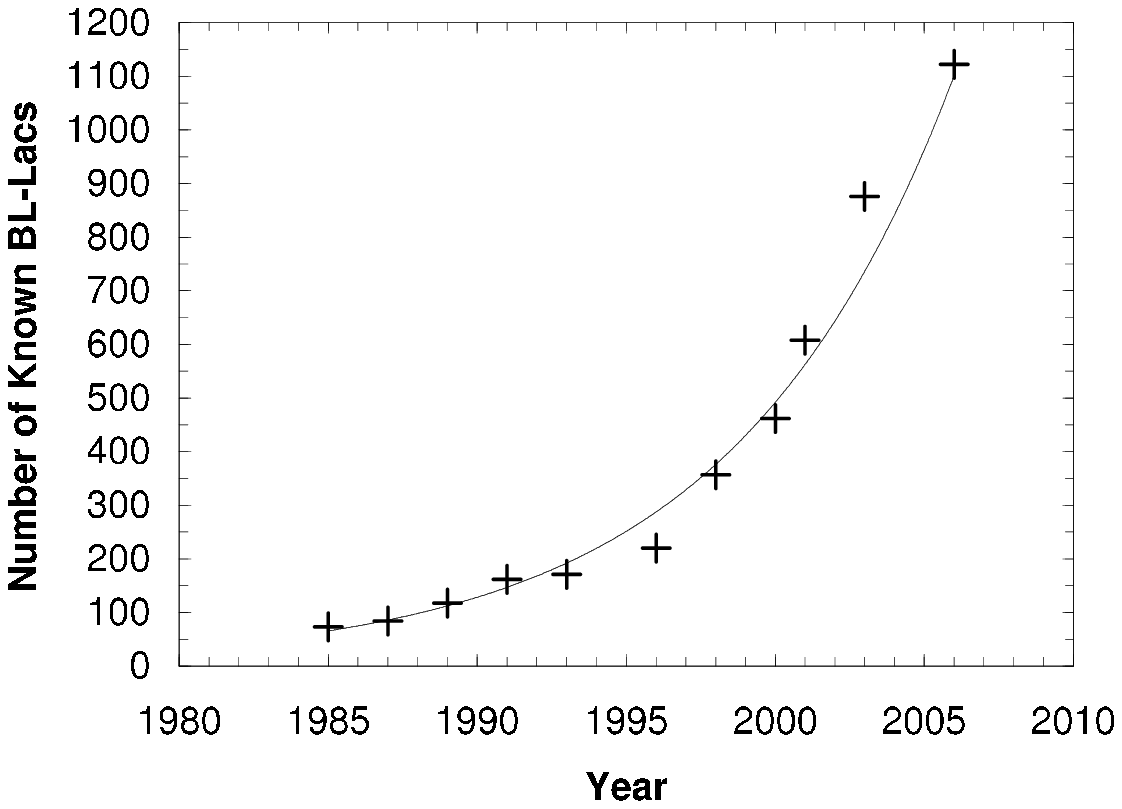}}
\figurecaption{1.}{Plot of the number of published \mbox{BL~Lac} in the
literature as a function of time, based on
Veron-Cetty \& Veron (1984, 1985, 1987, 1989, 1991, 1993, 1996, 
1998, 2000, 2001, 2003, 2006). For 2006
$\sim$1.2\% of all known AGNs are \mbox{BL~Lac},
compared to $\sim$88.8\% for QSO and
$\sim$10.0\% for Sy1. As new \textmd{\textit{ROSAT}}
\mbox{BL~Lac}s are discovery the number of known \mbox{BL~Lac}s
should exceed $\sim$2000 by 2010.}


Potentially, the largest complete catalogue of \mbox{BL~Lac} objects
remains mostly undiscovered in the \textit{ROSAT} All Sky Survey (RASS), a
fact first noted by Stocke~et~al.~(1989). Various studies have begun
the task of identifying RASS samples of \mbox{BL~Lac}s, the RGB sample (127 \mbox{BL~Lac}s) of
Laurent-Muehleisen~et~al.~(1999) and SDSS/RASS sample of Anderson~et~al.~(2007) with 
200 new BL~Lacs. The limiting factor for RASS sample sizes  
is the time consuming task of optical identification with a 15--20\arcsec\ position error. 

Presented in this paper is a sample of 18 new RASS \mbox{BL~Lac}s 
(including five (5) candidate objects and five (5) objects verified from the
literature) discovered as part of a radio detection survey and ESO
optical identification program for a complete sub-sample of 695 RASS sources.
Section~2 describes the RASS sub-sample selection, radio follow-up program and 
details how optical identifications were made and the
criteria used to select objects as \mbox{BL~Lac}s. In Section~3 we
describe the sample, including a table of \mbox{BL~Lac} properties.
In Section~4 we examine the properties of the sample and compare
them to samples from the literature. Finally, in Section~5 we
discuss the findings with Section~6 forming the conclusion.

\section{2. THE RASS SUB-SAMPLE}

The \textit{ROSAT} X-ray telescope completed the first All Sky 
Survey with an imaging X-ray telescope in 1991, resulting in two 
RASS Catalogues (unpublished RASS revision one catalogue, $\sim$54\,000 
sources, see Voges~(1993) and various RASS revision two 
catalogues, $\sim$160\,000 sources, see Voges~(1997) and Voges~et~al.~(1999, 2000)). 
To characterise the source population being detected a complete sub-sample of 
four (4) Key Fields, Table~1, was selected \mbox{($\vert$b$\vert$\,\texttt{>}--25\D,} $\delta$\texttt{<}10\D, 
695 sources, covering 575 square degrees) from the revision one catalogue 
for optical identification (Danziger~et~al.~1990; European Southern 
Observatory Key Field Program). To improve the multi-wavelength coverage and 
to assist the optical program all Key Field sources where observed at 4.8~GHz 
($\sim$1~mJy limit and $\sim$1$\arcsec$  positional accuracy). 

\vskip.5cm

\noindent {\bf Table 1.} Summary of the RASS Key Field properties.  

\vskip2mm 

{\small\centerline{\begin{tabular}{|c|c|c|c|} 
\hline \noalign{\smallskip}
 Key Field & RA & Dec & Number of Sources\\
\noalign{\smallskip}\hline\noalign{\smallskip}
I   & 01$^h$ & --40\D & 261\\
II  & 05$^h$ & --57\D & 227\\
III & 11$^h$ & --27\D & \p071 \\
IV  & 13$^h$ & +00\D  & 134\\
\noalign{\smallskip}\hline
\end{tabular}}

}

\vskip.5cm
 
The optical and radio follow-up identified a complete sample of 18~\mbox{BL~Lac}s using 
X-ray positions from the unpublished RASS revision one (1) catalogue. 
A positional correlation with the RASS revision two (2) catalogue establish all of the 
objects are real detections.

\subsection{2.1 Radio follow-up}

Radio observations of the \mbox{BL~Lac} sample (695) were undertaken as
part of the \mbox{4.8~GHz} radio detection program
(Anderson~2003; Anderson~et~al.~1994). Briefly, the Australia
Telescope Compact Array (ATCA) observed the RASS Key Field sources
in fields I, II and III, while the Very Large Array (VLA) observed sources in field IV.
For the ATCA all observations used a 1.5~km array configurations
plus antenna 6 and were made using a 128~MHz bandwidth divided into
32 channels of width 4~MHz. Closest match for the VLA is the C array
configuration, with a bandwidth of 50~MHz. All radio detections are
located within 2\arcmin\ of the pointing center where bandwidth
smearing effects are extremely small. The primary flux calibrators,
\mbox{PKS~1934-638} (ATCA) and \mbox{3C286} (VLA), are tied to the
VLA absolute flux scale of Baars~et~al.~(1977).

The radio data has been reduced using standard reduction procedures
within the Astronomical Image Processing Software (AIPS) package.
The observations reached a five sigma detection limit of
$\sim$2~mJy. All 4.8~GHz flux densities listed in
Table~2 have been primary beam corrected.

The radio positions were correlated with the NRAO VLA Sky Survey (NVSS) radio survey
(Condon~et~al.~1998) to determine \mbox{1.4 GHz} flux densities and errors for
\mbox{BL~Lac}s north of a declination of --42\D. Ten 
coincidences were found, with flux densities listed in
Table~2. The efficiency of combined radio (NVSS) and X-ray (RASS) can be 
as large as 35\%~(Beckmann~et~al.~2003). A similar comparison with the 
Sydney University Molonglo Sky Survey (SUMSS; Mauch~et~al.~2003) to determine 843~MHz 
flux densities found four source, listed in Table~2. 

Four \mbox{BL~Lac}s were verified by a positional correlation of
the Key Field radio positions with \mbox{BL~Lac} objects in the
NASA/IPAC Extragalactic Database (NED).

\subsection{2.2 Optical follow-up and identification}

Preliminary optical candidates were selected using optical finder
charts extracted from the COSMOS catalogue (Yentis~et~al.~1992).
Candidates were prioritised according to distance from the RASS
position and optical magnitude, with priority given to optical 
sources associated with a radio counterpart. Optical spectroscopy was then used
to identify one or more optical candidates until the most likely
identification was established. 

\mbox{X-ray} selected \mbox{BL~Lac}s are
\mbox{X-ray} loud and radio loud (Stocke~et~al.~1991). 
In practical terms this means all EMSS (Rector~et~al.~2001) and 
ESS (Perlman~et~al.~1996) BL~Lacs have a 
4.8~GHz flux density brighter than 1~mJy. Hence all
RASS \mbox{BL~Lac}s should be positionally associated with a radio counterpart
brighter than 1~mJy. If all BL~Lacs are required to be associated with a radio source then 
false identification from stellar source with featureless spectra would also be minimised. 
In the Key Field sample only one stellar X-ray sources (Anderson\& Filipovi\'c 2007) 
was detected as a radio emitter. 

\mbox{BL~Lac} object classification was based on the technique developed 
for the \textit{Einstein} Extended Medium Sensitivity Survey 
(Stocke~et~al.~1991). A source was classified as a \mbox{BL~Lac} if it had:
\item{(i)} A featureless optical spectrum i.e. no observed emission lines of equivalent width $\geq$0.5~nm.
\item{(ii)} A Ca \textsc{ii} break of less than 25\% to distinguish between a non-thermal AGN spectrum and the stellar spectrum of an elliptical galaxy.
\item{(iii)} A 4.8~GHz flux density greater than 1~mJy. 

Any candidate BL-Lac object was required to meet criteria three and have at-least a featureless spectrum.

More details about the optical program is described by
Danziger~et~al.~(1990). We recognise that the strength of the Calcium break is depending on
the luminosity of the object: the strength of the Calcium break is
anti-correlated with the luminosity (Landt~et~al.~2002; Beckmann~et~al.~2003). Hence, some
BL~Lac objects could exhibit Calcium breaks exceeding 25\% (Anderson~et~al.~2007). In addition,
blazars can show emission lines. As the BL~Lac objects are a
subclass of the blazar population and as there is apparently a
smooth transition from BL~Lacs to Flat Spectrum Radio Quasars (FSRQ), one has to be very careful
when applying strict constraints on the optical spectra. While 
we acknowledge that we might have missed border-line BL~Lac objects, 
the use of criteria three should help reduce the number missed.

\section{3. THE \mbox{BL~LAC} SAMPLE}

A total of 18 \mbox{BL~Lac}s have been identified, their
characteristics are summarised in Table~2;
\mbox{Column (1)} is the table index for cross referencing with the
index in Table~4; \mbox{Column (2)} is the
\textit{ROSAT} All Sky Survey Name; \mbox{Column (3)} is the COSMOS
J band magnitude; \mbox{Column (4)} is the redshift; \mbox{Column
(5)} is the radio position right ascension for J2000; \mbox{Column
(6)} is the radio position declination for J2000; \mbox{Column (7)}
is the ATCA \mbox{4.8~GHz} flux density and error in mJy; \mbox{Column (8)} is
the \mbox{1.4~GHz} flux density in mJy from either the Parkes 90
catalogue or the NVSS (only available for
sources north of a declination of --42\D) and associated errors; \mbox{Column (9)} is the
reference flag defined at the end of Table~2 and
indicates if a source is a candidate or confirmed Key Field
\mbox{BL~Lac} or is from the literature; \mbox{Column (10}) is the
RASS count rate and associated error and \mbox{column (11)} is the
Total Hydrogen Column Density in units of \mbox{10$^{20}$ atoms
cm$^{-2}$}. For additional details consult the source notes at the
end of this section.

A total hydrogen column density, N$_{\rm H}$, value (Col.~11) is needed
to correct the calculated \textit{ROSAT} Position Sensitive 
Proportional Counter (PSPC) flux for galactic
absorption. Values were estimated for Key Field sources north of a
declination of --42\D~using the Bell Labs \HI\ survey of
Stark~et~al.~(1992). For Key Field sources south of a declination of
--42\D, estimates were taken from the Parkes \HI\ survey of
Cleary~et~al.~(1979). Only values from Stark~et~al.~(1992) are not
affected by stray radiation entering the far field antenna pattern.

Thirteen (13) \mbox{BL~Lac}s, including five (5) candidates, have
been optically identified by Danziger~et~al.~(1990) and are listed in
Table~2. Up to June 2009, a positional correlation
of the radio positions with objects in the NASA/IPAC Extragalactic
Database (NED) identified five (5) additional sources classified in
the literature as \mbox{BL~Lac}s and listed in
Table~2.

Multi-epoch 4.8~GHz radio observations, based on a set of 40 sources,
indicates the flux density error is 
well represented by $ \sigma_{\rm R} = 0.08S_{\rm R}^{-1} + 0.48 $, with $S_{\rm R}$ 
the flux density in mJy over the range of  1.5~mJy to 1~Jy. Above 30~mJy 
the flux error is constant at about 8\%, rising to 40\% at 1.5~mJy, with 
errors for individual flux densities given in Table~2.  

The number of randomly associated radio counterparts for the entire
Key Field sample of 695 RASS objects is estimated at less than two,
based on 4.8~GHz radio source counts. This makes the chances of a
\mbox{BL~Lac} being a random association extremely small.

The Key Field sample is complete to approximately
\mbox{$(9\pm2)\times10^{-13}$\,erg\,cm$^{-2}$ s$^{-1}$}, at which, the
prior counts of Wolter~et~al.~(1991) predict a sky density of
0.030$\pm$0.008 per degree squared or 17$\pm$5
\mbox{BL~Lac}s over the 575 deg\SQR~of sky covered by the four Key
Fields. This value compares well to the observed value of
18$\pm$4 \mbox{BL~Lac}s observed, indicating a high level
of completeness. The number of \mbox{BL~Lac}s missed is estimated to
be less than two.

\subsection{3.1. Positional accuracy}

Astrometric accuracy of the radio and \mbox{X-ray} positions
determined by Anderson~(2003) are $1.1\arcsec$ and $30\arcsec$
respectively. For the radio positions the accuracy was determine by
comparing the positions of a standard set of 40 sources observed at
two or more epochs, giving a deviation of $\Delta\delta=0.8\arcsec$
and $\Delta\alpha cos(\delta)=0.7\arcsec$. RASS positional accuracy
was determined by comparing the \mbox{X-ray} and radio positions for
the $\sim120$ radio emitting \mbox{X-ray} sources identified by the
Key Field program, with $95\%$ of the radio sources located within
$40\arcsec$ of the \mbox{X-ray} position.

Comparison of the difference between the RASS and ATCA radio
positions for the 18 \mbox{BL~Lacs} in Fig.~2
indicates $90\%$ of identified sources are located within
$15\arcsec$. This value is smaller than the $30\arcsec$ error for
the entire Key Field sample and is probably due to \mbox{BL~Lac}
morphology being compact.

\centerline{\includegraphics[width=75mm]{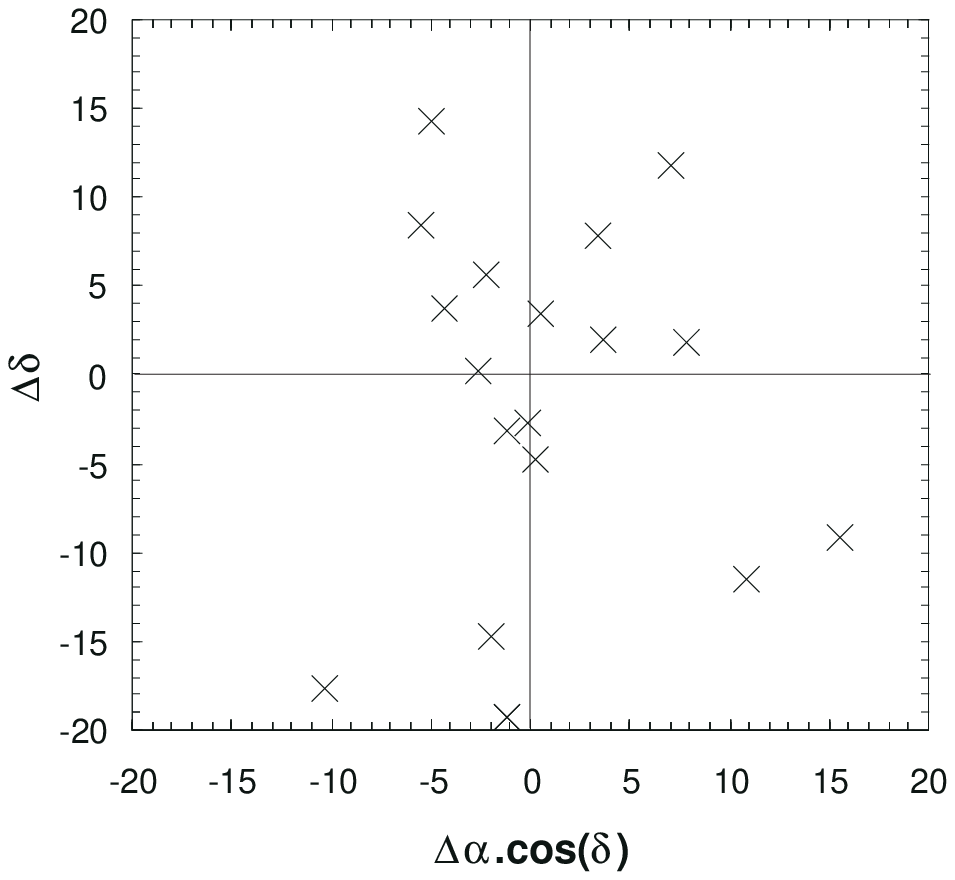}}
\figurecaption{2.}{The positional difference between the RASS \mbox{X-ray}
and the ATCA radio position. All \mbox{BL~Lac's} are
located within 20\arcsec~from the \mbox{X-ray} position.}

\subsection{3.2. Source notes}

\mbox{BL~Lac}s classified as candidate objects include the RASS
source \mbox{RX~J0049.7--4151}, \mbox{RX~J0045.8--3158}, \mbox{RX~J0059.5--3510}, 
\mbox{RX~J0054.8--2455} and \mbox{RX~J1357.2--0146}.

Objects confirmed as \mbox{BL~Lac}s by the ESO optical
identification program include \mbox{RX~J0109.9--4020}, 
\mbox{RX~J0140.9--4130}, \mbox{RX~J0543.9--5532}, \mbox{RX~J0506.9--5435} (also
known as \mbox{1ES~0505--546}), \mbox{RX~J0528.8--5920} (located next
to a field galaxy), \mbox{RX~J0528.8--5920}, \mbox{RX~J0439.0--5558},
\mbox{RX~J1357.6+0128} and \mbox{RX~J1103.6--2329}. Multi epoch radio
data for \mbox{RX~J0439.0--5558} reveals a variability of $28\%$ or
~3 mJy. \mbox{RX~J0109.9--4020} was independently confirmed as a
\mbox{BL~Lac} (see Schwope~et~al.~2000), with the radio source
\mbox{PMN~J0110--4020} one arcminute away. \mbox{RX~J0543.9--5532} was
independently identified in NED as a \mbox{BL~Lac} (see
Fischer~et~al.~1998). \mbox{RX~J1103.6--2329} was independently
confirmed by Laurent-Muehleisen~et~al.~(1993) and Perlman~et~al.~(1996)
as a \mbox{BL~Lac}, and is also known as \mbox{1ES~1101--232}, and
\mbox{PMN~J1101--2329} which has a 4.8~GHz flux density of 66$\pm$11~mJy 
that compares well to the ATCA value.

Five additional objects are classified as \mbox{BL~Lac}s in the
literature. Bauer~et~al.~(2000) identified \mbox{RX~J0040.3--2719} as a
\mbox{BL~Lac} and it is also coincident with the radio source
\mbox{NVSS J004016--271912}. \mbox{RX~J0043.4--2639} was confirmed as
\mbox{BL~Lac} by Cristiani~et~al.~(1995), also known as \mbox{PMN
J0043--2639} which has a slightly higher 4.8~GHz flux density of 81$\pm$11~mJy 
compared to the ATCA value 68$\pm$6~mJy. Bade~et~al.~(1994) confirmed \mbox{RX~J1057.8--2753} to be
a \mbox{BL~Lac} object. Veron-Cetty \& Veron (2006) confirmed 
\mbox{RX~J2324.7--4041} (also known as \mbox{1ES 2322--409}) as a \mbox{BL~Lac}.

Xue~et~al.~(2000) investigated \mbox{RX~J2319.1--4206}, showing it to
be a \mbox{BL~Lac} with a complex radio morphology and it is also
known as \mbox{PKS B2316--423} or \mbox{PMN J2319--4206}. The compact
core has a ATCA \mbox{4.8~GHz} flux density of \mbox{207 mJy}, while
the extended lobe has a 237~mJy flux density, giving a total of
444~mJy that compares well to the PMN single dish value of 595~mJy.
We also note that this object is very close to the galaxy cluster
\mbox{Abell~S1111} located 1.1\arcmin\ away. This would be indeed
very interesting, as BL~Lac objects usually seem to avoid cluster
environments (Wurtz~et~al.~1993, 1997). However, upon closers
inspection we found that the redshift of \mbox{RX~J2319.1--4206} is
z=0.055 and that of \mbox{Abell S1111} is z=0.045, which gives a
difference in distance of $\sim$50~Mpc. Thus, this BL~Lac is not
likely inside the galaxy cluster.

\section{4. PROPERTIES OF OUR SAMPLE}

The characteristics of the sample are examined by comparing it with
the \textit{Einstein} EMSS \mbox{BL~Lac} sample
(Gioia~et~al.~1990; Rector~et~al.~2001), the 1~Jy \mbox{BL~Lac}
sample (Stickel~et~al.~1991; Rector \& Stocke~2001) and the \textit{Einstein} Slew survey (ESS)
\mbox{BL~Lac} sample (Perlman~et~al.~1996). The 1~Jy sample is taken
to be representative of the properties of Low Energy Peaked
\mbox{BL~Lac}s (LBLs) and the EMSS of the properties of high
energy peaked \mbox{BL~Lac}s (HBLs). The ESS exhibits properties
that are intermediate between LBLs and HBLs. Comparison was made
of the distributions of $\nu_{\rm peak}$ (peak frequency of
the synchrotron emission), redshift, $\alpha_{\rm RO}$
versus $\alpha_{\rm OX}$, median spectral index and the
\mbox{X-ray} \mbox{BL~Lac} Log(N)--Log(S). The median values for the
four samples are given in Table~3.

\vskip.5cm

\noindent {\bf Table 3.} Median properties of the four \mbox{BL~Lac} samples. The median redshift
   value for this survey is an estimate as only 55\% of sources have redshifts.  

\vskip2mm 

{\small\centerline{\begin{tabular}{|l|c|c|c|c|} 
\hline \noalign{\smallskip}
                                  & EMSS  & 1 Jy  & ESS   & This  \\
                                  &       & Radio &       & Study \\
  \noalign{\smallskip}\hline\noalign{\smallskip}
Redshift                          & 0.30  & 0.55  & 0.16  & 0.31 \\
Log (S$_{{\rm X}}$/S$_{{\rm R}}$) & --4.8 & --6.9 & --4.6 & --5.0 \\
Log($\nu_{{\rm peak}}$)           & 16.2  & 13.8  & 15.4  & 15.9 \\
\noalign{\smallskip}\hline 
\end{tabular}}

}

\vskip.5cm

\subsection{4.1. Peak frequency of the synchrotron emission}
 \label{sect-peak-freq} 

The turnover frequency at which the spectral energy distribution
(SED) peaks can be used to distinguish LBLs from HBLs. Following
Sambruna~et~al.~(1996) and Landau~et~al.~(1986) the key field sample
flux densities for the radio (S$_{\rm R}$) optical (S$_{\rm O}$) and
\mbox{X-ray} (S$_{\rm X}$) were fitted with a logarithmic parabola of
the form:
\begin{equation}
Log\left( \nu S_{\nu } \right) =a(Log\left( \nu \right) )^{2} +bLog\left(
\nu \right) +c
\label{equation-one}
\end{equation}
with $\nu$ the observed frequency. The peak frequency is
then given by $Log\left( \nu _{\rm peak} \right) =-b/2a$ and was
k-corrected to the rest frame using $\nu _{\rm rest} =\nu \left(
1+z\right)$ were $\nu _{\rm rest}$ is the rest frame frequency. 

For the seven (7) \mbox{BL~Lac}s with no measured redshift, a value equal to
the sample median (z=0.31, Table~3) is used. 
While a median will likely underestimate the true redshift (i.e. 
nearby BL~lacs (z\texttt{<} 0.2) tend to have measured redshift due to the 
host galaxy being resolved, while objects with no measurable redshift 
tend to be at z\texttt{<}1) any associated error in $\alpha_{\rm RX}$ due to 
a K-correction would be smaller then 0.04 for z\texttt{<}1. 

Only radio, optical and \mbox{X-ray} data from Table~2 was
used. Flux densities were calculated for the PSPC count rate at 2~keV 
and for the {\rm COSMOS} optical magnitude at 641~THz using
S$_{\rm O}$=3.95$\times$10$^{-23}$ 10$^{-0.4m(J)}$ adopted
from Fukugita~et~al.~(1995) for S$_{\rm O}$ values calculated within the
NED database. An example of the SED and functional fit for one
source is given in Fig.~3. The \mbox{k-corrected}
peak values are listed in Table~4. A fit to source 7
\mbox{(RX~J0109.9--4020)} was problematic, giving a very high peak
frequency that has been placed in brackets in Table~4. 
Examination of the broad band fluxes shows the X-ray count rate is 
abnormally high (see Sect.~5).

\vskip.5cm
\centerline{\includegraphics[width=75mm]{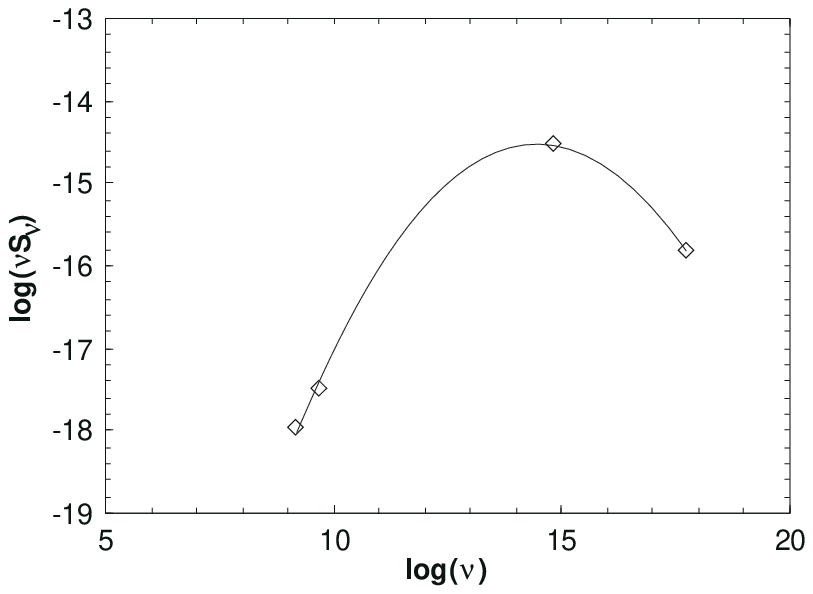}}
\figurecaption{3.}{Example of the logarithmic parabola fitted
(equation~1) to the spectral energy
distribution of source \mbox{RX~J0043.4--2639}, Table~4,
with the observed Log($\nu_{\rm peak}$)=14.81.}


Fig.~4 compares the two point radio to
\mbox{X-ray} spectral index, $\alpha_{\rm RX}$, versus
Log($\nu_{\rm peak}$) diagram for this sample
(Table~4) with that of the one (1) Jy Radio and the
EMSS using values from the literature (Sambruna~et~al.~1996). Clearly
$\alpha_{\rm RX}$ and Log($\nu_{\rm peak}$) are
correlated. A linear regression analysis of the Key Field data,
excluding one anomalous point (source 7 marked with $\ddag$ in
Table~2 and plotted as solid diamonds in
Fig.~4), results in:
\begin{displaymath}
\alpha _{\rm RX} =-0.091\times Log\left( \nu _{\rm peak} \right) +2.12
\end{displaymath}
In good agreement to the fit of Dong~et~al.~(2002),
\begin{displaymath}
\alpha _{\rm RX} =-0.09535\times Log\left( \nu _{\rm peak} \right)
+2.16065,
\end{displaymath}
based on the EMSS and 1~Jy radio samples.

Two lines defined by $\alpha_{\rm RX}$=0.75 and
Log($\nu_{\rm peak}$)=14.7 in
Fig.~4 can be used to delineate four
regions. HBLs occupy the lower right quadrant and LBLs the upper
left quadrant. This distinction was first suggested by
Dong~et~al.~(2002) and is equivalent to using
\mbox{Log(Sr/Sx)=--5.5} (Laurent-Muehleisen~et~al.~1999) as a criteria.
Intermediate \mbox{BL~Lac}s occupy the region defined by
$\alpha_{\rm RX}$=0.7 to $\alpha_{\rm RX}$=0.8.

The Key Field sample is composed mostly of 13 HBLs, three IBLs and
two others falling within the region for LBLs. A median
\mbox{Log($\nu_{\rm peak}$)=15.9} (Table~3)
is close to the EMSS value of 16.2, both are typical of HBLs. As
expected the ESS with IBLs has a median value
\mbox{Log($\nu_{\rm peak}$)=15.4} that falls close to
halfway between the LBLs (1 Jy sample)
\mbox{Log($\nu_{\rm peak}$)=13.8} and the HBLs
\mbox{Log($\nu_{\rm peak}$)=16.2} (EMSS).

\vskip.5cm

\noindent {\bf Table 4.} Values of the spectral indices for radio to optical, optical to X$-$ray
   and radio to X$-$ray, the logarithm of the frequency for the SED turnover
   peak and the type of BL~Lac, LBL, IBL and HBL. Note that Col.~1 (Index) is for cross 
   referencing with Table~1.  

\vskip2mm 

{\small\centerline{\begin{tabular}{|c|c|c|c|c|l|} 
\hline \noalign{\smallskip}
Index & $\alpha_{{\rm RO}}$ & $\alpha_{{\rm OX}}$ & $\alpha_{{\rm RX}}$ & Log($\nu_{{\rm peak}}$) & Type \\ 
\noalign{\smallskip}\hline\noalign{\smallskip}
 1 & 0.42 & 1.12 & 0.70 & 15.9 & HBL \\
 2 & 0.41 & 1.52 & 0.83 & 14.8 & LBL \\
 3 & 0.38 & 0.99 & 0.63 & 16.6 & HBL \\
 4 & 0.42 & 1.09 & 0.69 & 16.0 & HBL \\
 5 & 0.35 & 1.22 & 0.68 & 15.5 & HBL \\
 6 & 0.50 & 1.03 & 0.72 & 16.4 & IBL \\
 7 & 0.56 & 0.66 & 0.64 & (19.2) & HBL \\
 8 & 0.15 & 1.37 & 0.60 & 15.3 & HBL \\
 9 & 0.37 & 1.03 & 0.64 & 16.3 & HBL \\
10 & 0.28 & 0.94 & 0.55 & 16.8 & HBL \\
11 & 0.46 & 1.17 & 0.74 & 15.8 & IBL \\
12 & 0.32 & 0.94 & 0.57 & 16.8 & HBL \\
13 & 0.44 & 1.07 & 0.69 & 16.0 & HBL \\
14 & 0.32 & 0.90 & 0.56 & 16.9 & HBL \\
15 & 0.36 & 1.14 & 0.67 & 15.9 & HBL \\
16 & 0.37 & 1.33 & 0.73 & 15.2 & IBL \\
17 & 0.33 & 1.66 & 0.82 & 14.4 & LBL \\
18 & 0.22 & 1.17 & 0.58 & 15.8 & HBL \\
\noalign{\smallskip}\hline
\end{tabular}}

}

\vskip.5cm

\subsection{4.2. Redshift distribution}

The Key Field sample has 11 \mbox{BL~Lac}s with measured redshifts 
(eight (8) have z\texttt{<}0.5 and three (3) have z\texttt{>}0.5) 
and seven with no redshift. Six of the seven with no redshift are
bright (i.e m$_{\rm J}$\texttt{<}18) and have \mbox{Log(S$_{\rm m(J)}/S_{\rm X}$)}
typical of objects in the EMSS with a z\texttt{<}0.8
(Stocke~et~al.~1991). The median redshift of the sample (z=0.31) is 
consistent with the EMSS (z=0.30) and is placed about halfway 
between the median value for the ESS (z=0.16) and 1~Jy (z=0.55) samples.
Similarly the median z and Log(S$_{\rm X}$/S$_{\rm R}$) for the three samples and the Key
Field value of \mbox{Log(S$_{\rm X}$/S$_{\rm R}$)=--5.0} corresponds to an
interpolated median z of 0.35$\pm$0.4. 

The Key Field sample indicates a sample of HBLs has a
systematically lower redshifts than a LBLs (i.e. 1~Jy) sample.
This finding is consistent with the unification model of
Fossati~et~al.~(1997), which predicts HBLs to have systematically
lower redshifts compared to LBLs. Meaning that HBLs have lower
intrinsic luminosities than LBLs. A slight correlation was found
for the Key Field and EMSS sample between z and
Log($\nu_{\rm peak}$), indicating
Log($\nu_{\rm peak}$) is increasing with z. The same
variables do not correlate for the 1~Jy sample.

\end{multicols}

\centerline{\includegraphics[width=0.85\textwidth]{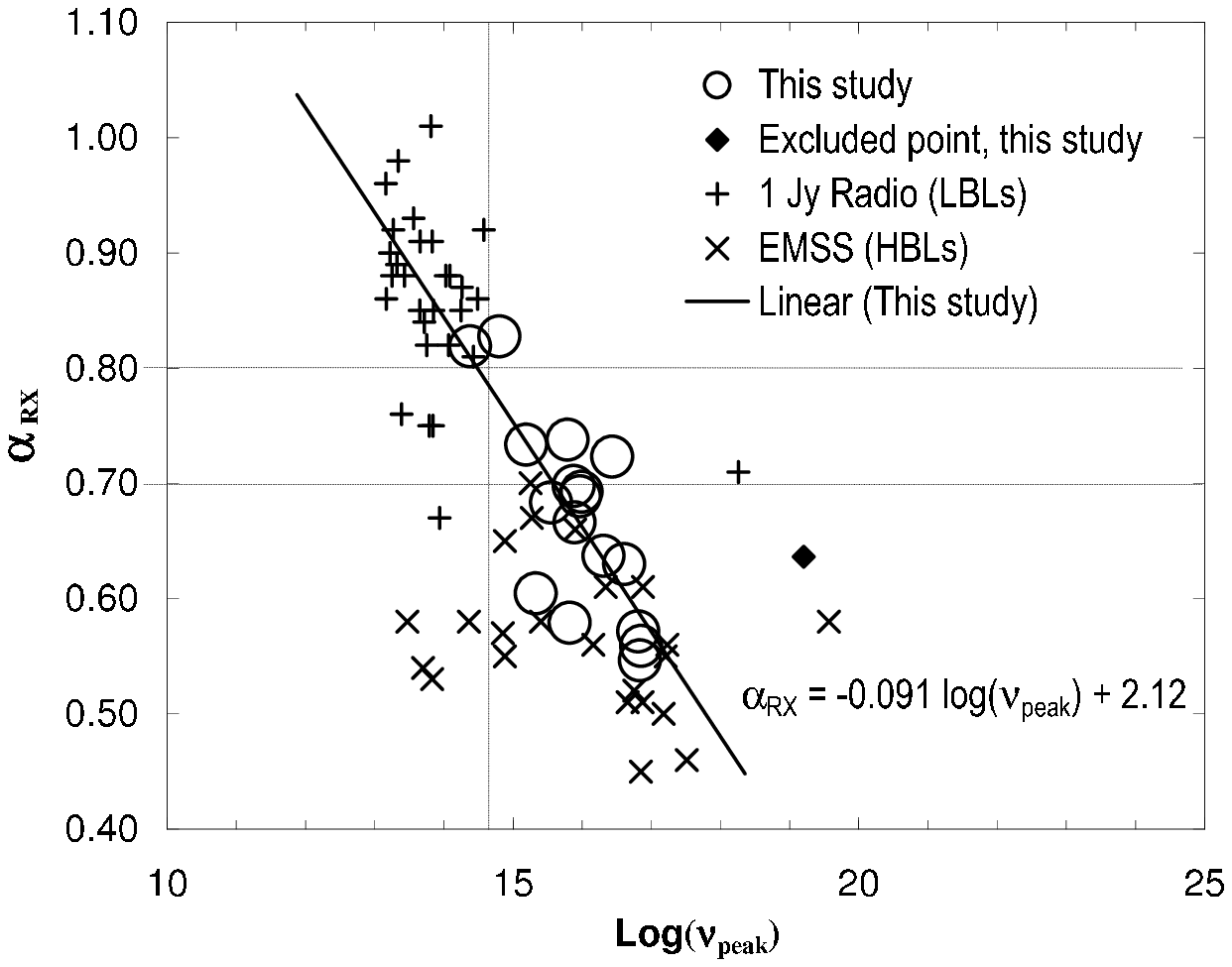}}
\figurecaption{4.}{The positional difference between the RASS \mbox{X-ray}
position and the ATCA radio position. All \mbox{BL~Lac's} are
located within 20\arcsec~from the \mbox{X-ray} position.}
\label{fig3-aRX-log-v-peak}

\centerline{\includegraphics[width=0.85\textwidth]{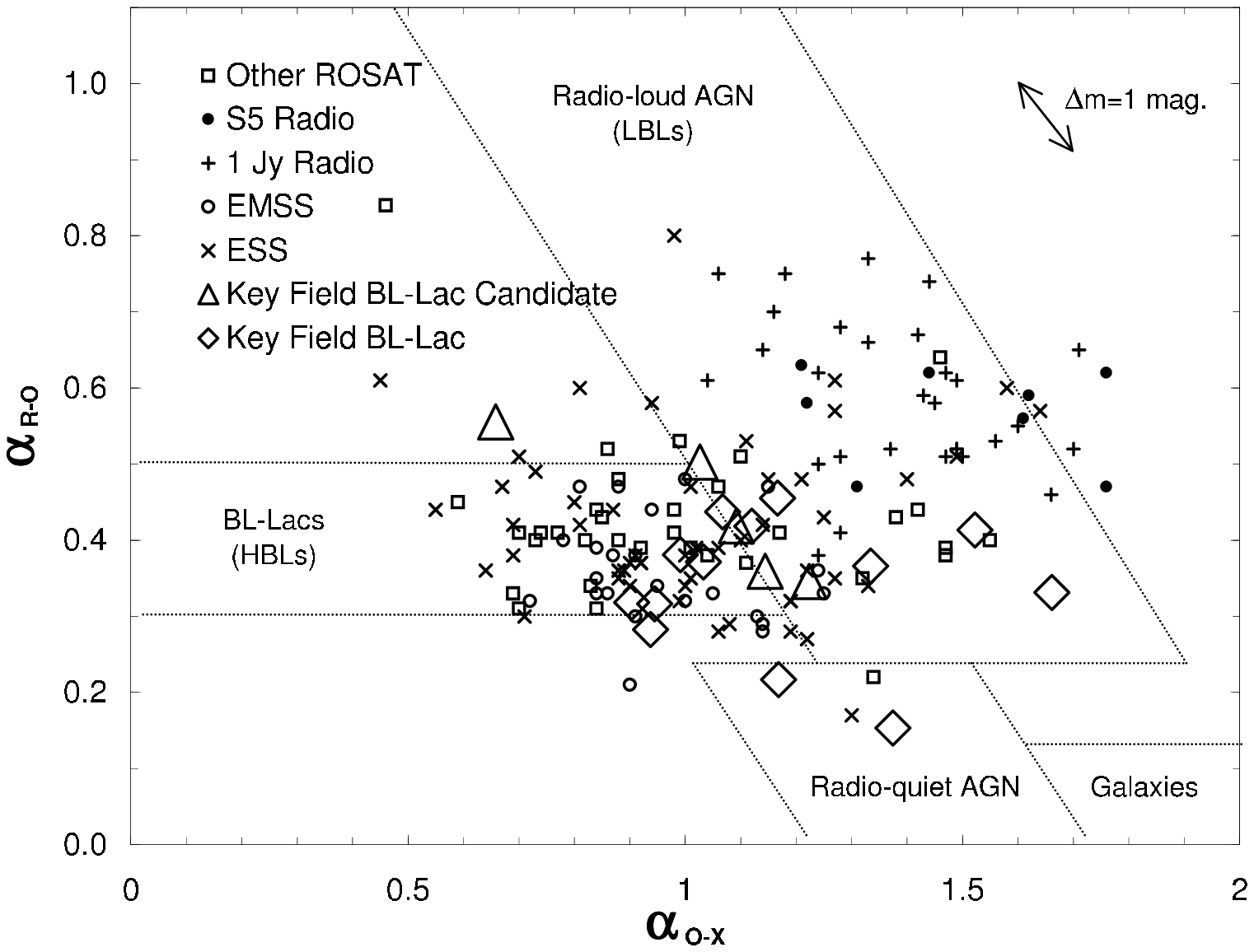}}
\figurecaption{5.}{Radio to optical versus optical to \mbox{X-ray} spectral
index plot for \mbox{BL~Lac} objects. The Key Field sample,
including candidate objects, are plotted as a large diamond and
triangle, respectively. Other samples shown are for comparison and
are drawn from \mbox{X-ray} selected samples; \textit{Einstein}
EMSS (Giommi~et~al.~1995), \textit{Einstein} Slew survey (ESS;
Perlman~et~al.~1996), other \textit{ROSAT} \mbox{BL~Lac}s and
radio selected \mbox{BL~Lac}s; 1~Jy and S5 (Stickel~et~al.~1991). The
regions defining different types of objects, and the definition of
$\alpha_{\rm RO}$ and $\alpha_{\rm OX}$, are taken
from the \textit{Einstein} EMSS (Stocke~et~al.~1991). A one
magnitude optical variation is indicated in the top right corner.}
\label{fig4-2colour-index}

\begin{multicols}{2}
{

\subsection{4.3. Spectral index}

The method of Brinkmann\& Siebert (1994) was used to estimate a spectral
index for each \mbox{BL~Lac} using hardness ratios. The hardness
ratios for each \mbox{BL~Lac} have large errors that make the
estimated spectral index values uncertain. Therefore, only the
sample median value will be presented. For the 18 sources the median
energy index is 1.42$\pm$0.80 and compares well to values
from the literature, 1.32$\pm$0.40 for a RASS sample
(Laurent-Muehleisen~et~al.~1999) and 1.47$\pm$0.40 for the
EMSS (Rector~et~al.~2001).

\subsection{4.4. $\alpha_{\rm RO}$ versus $\alpha_{\rm OX}$ diagram}

Fig.~5 is the $\alpha_{RO}$ versus
$\alpha_{OX}$ plot for the Key Field and comparison samples. The two
point spectral index is calculated using
$\alpha_{\rm RO}$=Log(S$_{\rm R}$/S$_{\rm O}$)/Log($\nu_{\rm R}$/$\nu_{\rm O}$)
and
$\alpha_{\rm OX}$=Log(S$_{\rm O}$/S$_{\rm X}$)/Log($\nu_{\rm O}$/$\nu_{\rm X}$)
with flux densities calculated at \mbox{$\nu_{\rm R}$=4.8~GHz,}
\mbox{$\nu _{\rm O}$=641~THz} ({\rm COSMOS} III-aJ Photographic B
magnitude is $\sim$ B $\lambda$=4400 {\AA},
see Sect.~4.1 for further details) and
$\nu_{\rm X}$ = 5.40$\times$10$^{17}$ Hz
(\mbox{X-ray} flux density is calculated for 2 keV). The flux
densities have been k corrected to the rest frame (i.e.
S$_{\rm rest}$=S$\times$(1+z)$^{\alpha \rm -1}$),
assuming $\alpha_{\rm R}$=0.0,
$\alpha_{\rm O}$=1.0 and $\alpha_{\rm X}$=1.2),
with the median sample redshift (Table~3) used if
none are available. The final values are listed in
Table~4, along with $\alpha_{\rm RX}$.
Comparison values are based on data in the literature.

The one anomalous $\nu_{\rm peak}$ value from
Sect.~4.1 corresponds to an extreme outlier in
the $\alpha_{\rm RO}$ versus $\alpha_{\rm OX}$
plane above the main \mbox{BL~Lac} region. The value is flagged in
Table~4 by brackets and is likely unrepresentative
of the overall distribution (see Sect.~5).

The other 15-points (4 candidate and 11 confirmed), likely HBLs, are positioned
within the regions occupied by the EMSS and the \mbox{1~Jy} samples. Two
points, likely IBLs, are located in the radio quiet region, but are still within
the region occupied by ESS \mbox{BL~Lac}s.

\subsection{4.5. X$-$ray Log (N) -- Log (S)}

The X-ray source counts (Fig.~6) are 
consistent with the EMSS and the ESS counts. They have
a slight bump at \mbox{f$_{\rm X}$=10$^{-11}$ erg cm$^{-2}$ s$^{-1}$}
reported by Maccacaro~et~al.~(1989) and Nass~et~al.~(1996). The EMSS
counts fall well below the Key Field level as does the ESS count.
This is partially due to incompleteness of the EMSS to bright
\mbox{X-ray} sources and the ESS to faint \mbox{X-ray} sources
between \mbox{$\sim$2$\times$10$^{-12}$ erg
cm$^{-2}$ s$^{-1}$} and \mbox{10$^{-11}$ erg cm$^{-2}$ s$^{-1}$.} 

\vskip0.5cm
\centerline{\includegraphics[width=75mm]{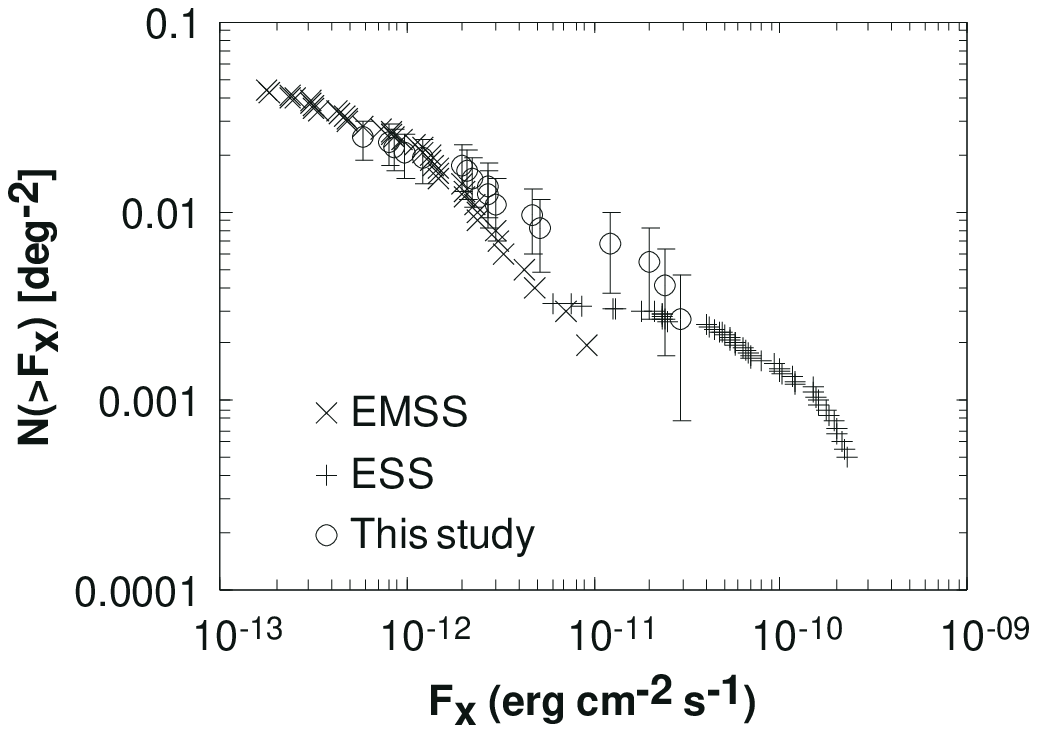}}
\figurecaption{6.}{X-ray Log (N)-Log(S) for the Key Field \mbox{BL~Lac}s and
comparison samples. All error bars are based on the count error.}
\label{fig5-x-ray-source-counts}

Taking the flux limit of the RASS as
\mbox{$\sim$3$\times$10$^{-13}$ erg cm$^{-2}$
s$^{-1}$} (0.1-2.4 keV) corresponds to a sky density of 0.04
\mbox{BL~Lac}s per square degree. At this density the RASS will
contain \mbox{(2.6$\pm$0.4)$\times$10$^{3}$}
\mbox{BL~Lac}s for the whole sky, reducing to
\mbox{(2.0$\pm$0.4)$\times$10$^{3}$} at high galactic
latitudes.

\section{DISCUSSION}

A candidate radio quiet BL~Lac, RX~J0140.9--4130, with a 1.8~mJy flux density 
occupies the radio quiet AGN region of the $\alpha_{\rm RO}$--$\alpha_{\rm OX}$ 
plot at $\alpha_{\rm RO}$=0.15 (Fig.~5). A total of 16 BL~lacs in the sample have 
$\alpha_{\rm RO}>0.3$, more than double the 0.15 value.
This contrasts to most X-ray selected samples that tend to be
\mbox{X-ray} loud and radio loud (Stocke~et~al.~1991).

The unusual source, RX~J0109.9--4020, has a SED with a very high synchrotron peak frequency with 
$Log\left( \nu _{\rm peak} \right)\approx19.2$. Using all available flux measurements 
we looked for possible abnormal flux values that might explain the high 
$Log\left( \nu _{\rm peak} \right)$ value. 
No indication of variability was found, while using only fluxes close to the 
X-ray flux epoch (1991) does not alter the $Log\left( \nu _{\rm peak} \right)$ value.  
The only noticeable difference is the RASS flux is significantly high relative 
to the other optical and X-ray fluxes i.e. $\alpha_{{\rm OX}}=0.66$ 
while $\alpha_{{\rm OX}}>1$  for the 17 other objects. Alternatively, RX~J0109.9--4020 
could be an example of a new population of BL~Lacs, possibly a ultra high energy  
peaked BL~lac predicted by Ghisellini~(1999) to have a 
peak frequency $\nu _{\rm peak}$ located at higher frequencies than HBLs, 
$Log(\nu _{\rm peak})>18$. Nieppola~et~al.~(2006) and Wu~et~al.~(2009) showed for a large BL~Lac sample
their is a population of extreme HBL with $Log(\nu _{\rm peak})$ values extending out to 21.5.

Traditionally the $\alpha_{\rm RO}$--$\alpha_{\rm OX}$ plot has been 
used to demonstrate the divisions of the population. These boundaries have 
proven successful in identifying new \mbox{BL~Lac} objects, achieving efficiencies of
$\sim$20\% (e.g. \textit{Einstein} Slew Survey (ESS);
Perlman~et~al.~1996; Schachter~et~al.~1993). If the EMSS colour-colour
boundaries had been strictly applied to the Key Field BL~Lacs then up to seven 
BL~Lacs would have been missed. 
 
A problem facing \mbox{BL~Lac} samples is the use of selection
criteria and the associated biases introduced. An example of this is the
observed bimodality between LBLs (1~Jy) and HBLs (Key Field and
EMSS) which is partly due to \mbox{BL~Lac} selection, in that the
1~Jy and EMSS/Key-field samples contain objects biased to the
selection of \mbox{BL~Lac}s using radio or \mbox{X-ray} properties
only. Laurent-Muehleisen~et~al.~(1999) showed the \textit{ROSAT} Green
bank \mbox{BL~Lac} sample has no bimodality owing to a large number
of intermediate \mbox{BL~Lac}s. 

By cutting off the Key Field sample to
simulate an \mbox{X-ray} and Radio selected sample should reproduced the observed
bimodality. The EMSS limit in the {\it ROSAT} band is 
about $3\times10^{-2}$~$\mu$Jy (Laurent-Muehleisen~et~al.~1999), giving a median 
\mbox{Log(S$_{\rm X}$/S$_{\rm R}$)=--4.8}, which is exactly the 
same as the EMSS (HBL) value, Table~3. The only bright radio source in the sample was taken to be representative of the 1~Jy sample, giving \mbox{Log(S$_{\rm X}$/S$_{\rm R}$)}= --6.7 for the total flux, which is the same as the 1~Jy (LBL) median value of --6.9. Clearly even the Key Field sample can reproduce the observed
bimodality between HBL and LBL.

\section{CONCLUSION}

The primary result of this paper is the sample of 18 \mbox{BL~Lac}s 
presented in Table~2. Based on the sample properties (Table~4)  
their are two LBL, three IBLs and 13 HBLs, making HBLs 
over represented compared to LBLs that make up only 11\% of 
objects, as expected for a primarily X-ray selected sample. 
By applying appropriate X-ray and radio flux limits to the sample 
we where able to reproduce the observed bimodality between LBLs (1~Jy) and
HBLs (EMSS). Two unusual source have been identified, a candidate radio 
quiet BL~Lac, RX~J0140.9--4130, and an extreme HBL, RX~J0109.9--4020, with a  
very high synchrotron peak frequency \mbox{$Log\left( \nu _{\rm peak} \right)\approx19.2$}.
A slight positive correlation was found between $\nu_{\rm peak}$ and
redshift.

A correlation between $\alpha_{\rm RX}$ and
Log($\nu_{\rm peak}$) was also found defined by the fit
\mbox{$\alpha_{\rm RX} =-0.091Log\left( \nu _{\rm peak} \right) +2.12$} and is
in good agreement to the fit of Dong~et~al.~(2002). The fit can be
used to directly calculate $\nu_{\rm peak}$ when no optical
flux is available.

Two main problems are faced by \mbox{BL~Lac} studies, understanding the SED and how it relates to 
the properties of the underlying source population, and the small number of known objects. 
A number of major programs are constructing large BL~Lac samples ($\approx500$) by 
combining the RASS (Anderson~et~al.~2007) or FIRST (Plotkin~et~al.~2008) surveys 
with the SDSS for identification. At a flux limit of
\mbox{$\sim$3$\times$10$^{-13}$ erg cm$^{-2}$
s$^{-1}$ (0.1--2.4 keV)} the RASS contains $\sim$2592
\mbox{BL~Lac}s for the whole sky, reducing to $\sim$2000
at high galactic latitudes. Even a complete sample of only 500 objects would allow major
progress to be made in understanding the \mbox{BL~Lac} phenomena.
Our concept of existing \mbox{BL~Lac} objects based on radio or
\mbox{X-ray} selection may need to be modified to fit into
potentially new \mbox{BL~Lac} populations suggested by optical
selections Londish~et~al.~(2002) and Collinge~et~al.~(2005).


\acknowledgements{
We would like to thank John Danziger for providing valuable input in
optical identification of our 18 \mbox{BL~Lac} objects presented in
this paper. This research was supported by the Australian Telescope
National Facility (ATNF). This research has made use of NASA's
Astrophysics Data System: we have also used data obtained from the
High Energy Astrophysics Science Archive Research Center (HEASARC),
which is provided by NASA's Goddard Space Flight Center. We thank the 
referee for his/her excellent comments that have greatly improved this 
manuscript.
                   }

\references


    Anderson, M. W. B.:
    2003, A Radio Survey Of Selected Fields From The ROSAT All Sky Survey,
    Ph. D. Thesis, University of Western Sydney

	Anderson, M. W. B., Filipovi\'c, M. D.:
    2007, \journal{Mon. Not. R. Astron. Soc.}, \vol{381}, 1027

    Anderson, M. W. B., White, G. L., Ekers, R. D., Danziger, J.:
    1994, The First Stromlo Symposium: The Physics of Active Galaxies.
    ASP Conference Series, \vol{54}, 355

    Anderson, S. F., Margon, B., Voges, W., et~al.:  
    2007, \journal{AJ}, \vol{133}, 313

    Bade, N., Fink, H. H., Engels, D.:
    1994, \journal{A\&A}, \vol{286}, 381

    Baars, J. W. M., Genzel, R., Pauliny-Toth, I. I. K., Witzel, A.: 
    1977, \journal{A\&A}, \vol{61}, 99

    Bauer, F. E., Condon, J. J., Thuan, T. X., Broderick, J. J.:
    2000, \journal{ApJS}, \vol{129}, 547

    Beckmann, V., Engels, D., Bade, N., Wucknitz, O.:
    2003, \journal{A\&A}, \vol{401}, 927

    Brinkmann, W., Siebert, J.:
    1994, \journal{A\&A}, \vol{285}, 812

    Cleary, M. N., Haslam, C. G. T., Heiles, C.:
    1979, \journal{A\&AS}, \vol{36}, 95

    Condon, J. J., Cotton, W. D., Greisen, E. W., Yin, Q. F., Perley,
    R. A., Taylor, G. B., Broderick, J. J.:
    1998, \journal{AJ}, \vol{115}, 1693
    
    Collinge, M. J., Strauss, M. A., Hall, P. B., et~al.:
    2005, \journal{AJ}, \vol{129}, 2542

    Cristiani, S., La Franca, F., Andreani, P., et~al.:
    1995, \journal{A\&AS}, \vol{112}, 347

    Danziger, I. J., Tr\"umper, J., Beuermann, K., et~al.:
    1990, \journal{ESO Messenger}, \vol{62}, 4

    Dong, Y-M., Mei, D-C., Liang, E-W.:
    2002, \journal{PASJ}, \vol{54}, 171

    Fischer, J.-U., Hasinger, G., Schwope, A. D., Brunner, H.,
    Boller, T., Tr\"umper, J., Voges, W., Neizvestny, S.:
    1998, \journal{AN}, \vol{319}, 347

    Fossati, G., Celotti, A., Ghisellini, G., Maraschi, L.:
    1997, \journal{Mon. Not. R. Astron. Soc.}, \vol{289}, 136

    Fukugita, M.,  Shimasaku, K.,  Ichikawa, T.:
    1995, \journal{PASP}, \vol{107}, 945

    Ghisellini, G.:
    1999, \journal{ApL\&C}, \vol{39}, 17

    Gioia, I. M., Maccacaro, T., Schild, R. E., Wolter, A., Stocke,
    J. T., Morris, S. L., Henry, J. P.:
    1990, \journal{ApJS}, \vol{72}, 567

    Giommi, P., Ansari, S.~G., Micol, A.:
    1995, \journal{A\&AS}, \vol{109}, 267

    Giommi P., Padovani P.:
    1994, \journal{Mon. Not. R. Astron. Soc.}, \vol{268}, L51

    Kollgaard, R. I.:
    1994, \journal{Vistas in Astronomy}, \vol{38}, 29

    Landau, R., Golisch, B., Jones, T. W., et~al.:
    1986, \journal{ApJ}, \vol{308}, 78

    Landt, H., Padovani, P., Giommi, P.:
    2002, \journal{Mon. Not. R. Astron. Soc.}, \vol{336}, 945

    Laurent-Muehleisen, S. A., Kollgaard, R. I., Moellenbrock, G. A., Feigelson, E. D.:
    1993, \journal{AJ}, \vol{106}, 875

    Laurent-Muehleisen, S. A., Kollgaard, R. I., Feigelson, E. D., Brinkmann, W., Siebert, J.:
    1999, \journal{ApJ}, \vol{525}, 127

    Londish, D., Croom, S. M., Boyle, B. J., et~al.:
    2002, \journal{Mon. Not. R. Astron. Soc.}, \vol{334}, 941

    Maccacaro, T., Gioia, I. M., Schild, R. E., Wolter, A., Morris, S. L., Stocke, J. T.:
    1989, BL Lac Objects, Ed:, L. Maraschi, T. Maccacaro, M.-H. Ulrich, Springer-Verlag, 222

    Maccacaro, T., Caccianiga, A., della Ceca, A., Wolter, A., Gioia, I. M.:
    1998, \journal{AN}, \vol{319}, 15

    Mauch, T., Murphy, T., Buttery, H. J., Curran, J., Hunstead, R. W.,
	Piestrzynski, B., Robertson, J. G., Sadler, E. M.:
    2003, \journal{Mon. Not. R. Astron. Soc.}, \vol{342}, 1117

    Nass, P., Bade, N., Kollgaard, R. I., Laurent-Muehleisen, S. A.,
    Reimers, D., Voges, W.:
    1996, \journal{A\&A}, \vol{309}, 419

	Nieppola, E., Tornikoski, M., Valtaoja, E.:
    2006, \journal{A\&A}, \vol{445}, 441

    Padovani, P., Giommi, P.:
    1995, \journal{ApJ}, \vol{444}, 567

    Perlman, E. S., Stocke, J. T., Schachter, J. F., et~al.:
    1996, \journal{ApJS}, \vol{104}, 251

    Perlman, E. S., Padovani, P., Giommi, P., Sambruna, R.,
    Jones, L. R., Tzioumis, A., Reynolds, J.:
    1998, \journal{AJ}, \vol{115}, 1253

    Plotkin, R. M, Anderson, S. F., Patrick B. Hall, P. B., Margon, B., Voges, W., 
	Schneider, D. P., Stinson, G., York, D. G.:
    2008, \journal{AJ}, \vol{135}, 2453

    Prandoni, I., Gregorini, L., Parma, P., de Ruiter, H. R., Vettolani, G., Wieringa, M. H., Ekers, R. D.:
    2000,\journal{A\&AS}, \vol{146}, 41

    Rector, T. A., Stocke, J. T.:
    2001, \journal{AJ}, \vol{122}, 565

    Rector, T. A., Stocke, J. T., Perlman, E. S., Morris, S. L., Gioia, I. M.:
    2001, \journal{AJ}, \vol{120}, 1626

    Rector, T. A., Gabuzda, D. C., Stocke, J. T.:
    2003, \journal{AJ}, \vol{125}, 1060

    Sambruna, R. M., Maraschi, L., Urry, C. M.:
    1996, \journal{ApJ}, \vol{463}, 444

    Schachter, J. F., Stocke, J. T., Perlman, E., et~al.:
    1993, \journal{ApJ}, \vol{412}, 541

    Schwope, A., Hasinger, G., Lehmann, I., et~al.:
    2000, \journal{AN}, \vol{321}, 1

    Stark, A. A., Gammie, C. F., Wilson, R. W., Bally, J., Linke, R. A.,
    Heiles, C., Hurwitz, M.:
    1992, \journal{ApJS}, \vol{79}, 77

    Stickel, M., Fried, J. W., Kuehr, H., Padovani, P., Urry, C. M.:
    1991, \journal{ApJ}, \vol{374}, 431

    Stocke, J. T., Morris, S. L., Gioia, I. M., Maccacaro, T., Schild, R. E., Wolter, A.:
    1989, BL~Lac Objects: Proceedings of a workshop held in Como, Italy, September 20-23,
    1988, Springer-Verlag, 242

    Stocke, J. T., Morris, S. L., Gioia, I. M., Maccacaro, T., Schild, R., Wolter, A.,
    Fleming, T. A., Henry, J. P.:
    1991, \journal{ApJS}, \vol{76}, 813

    Urry, C. M., Padovani, P.:
    1995, \journal{PASP}, \vol{107}, 803

    Veron-Cetty, M. P., Veron, P.:
    A catalogue of quasars and active nuclei (1th Edition), 1984, ESO, 1

    Veron-Cetty, M. P., Veron, P.:
    A catalogue of quasars and active nuclei (2nd Edition), 1985, ESO, 1

    Veron-Cetty, M. P., Veron, P.:
    A catalogue of quasars and active nuclei (3rd Edition), 1987, ESO, 1

    Veron-Cetty, M. P., Veron, P.:
    A catalogue of quasars and active nuclei (4th Edition), 1989, ESO, 1

    Veron-Cetty, M. P., Veron, P.:
    A catalogue of quasars and active nuclei (5th Edition), 1991, ESO, 1

    Veron-Cetty, M. P., Veron, P.:
    A catalogue of quasars and active nuclei (6th Edition), 1993, \journal{ESO Sci. Rep.}, \vol{13}, 1

    Veron-Cetty, M. P., Veron, P.:
    A catalogue of quasars and active nuclei (7th Edition), 1996, \journal{ESO Sci. Rep.}, \vol{17}, 1

    Veron-Cetty, M. P., Veron, P.:
    A catalogue of quasars and active nuclei (8th Edition), 1998, \journal{ESO Sci. Rep.}, \vol{18}, 1

    Veron-Cetty, M. P., Veron, P.:
    A catalogue of quasars and active nuclei (9th Edition), 2000, \journal{ESO Sci. Rep.}, \vol{19}, 1

    Veron-Cetty, M. P., Veron, P.:
    A catalogue of quasars and active nuclei (10th Edition), 2001, \journal{ESO Sci. Rep.}, \vol{20}, 1

    Veron-Cetty, M. P., Veron, P.:
    2003, \journal{A\&A}, \vol{412}, 399

    Veron-Cetty, M. P., Veron, P.:
    2006, \journal{A\&A}, \vol{455}, 773

    Voges, W.:
    1993, \journal{Adv. Space Res.}, \vol{13}, 391

    Voges W.:
    1997, Proc. 5th workshop Data Analysis in Astronomy,
    Eds. V. Di Gesu, M. J. B., Duff, A. Heck, M. C. Maccarone,
    L. Scarsi, H. U. Zimmermann, World Sci. Pub. Co., 189

    Voges, W., Aschenbach, B., Boller, Th., Bruninger, H., et~al.:
    1999, \journal{A\&A}, \vol{349}, 389
 
    Voges, W., Aschenbach, B., Boller, Th., Bruninger, H., et~al.:
    2000, \journal{IAUC}, \vol{7432}, 1

    Wolter A., Gioia, I. M., Maccacaro, T., Morris, S. L., Stocke, J. T.:
    1991, \journal{ApJ}, \vol{369}, 314

    Wolter A., Ciliegi P., Della Ceca R., Gioia I. M., Giommi P.,
    Henry J. P., Maccacaro T., Padovani P., Ruscica C.:
    1997, \journal{Mon. Not. R. Astron. Soc.}, \vol{284}, 225

    Wu, Z., Gu, M., Jiang, D. R.:
    2009, \journal{Research in Astronomy \& Astrophysics}, \vol{9}, 168

    Wurtz, R., Ellingson, E., Stocke, J. T., Yee, H. K. C.:
    1993, \journal{AJ}, \vol{106}, 869

    Wurtz, R., Ellingson, E., Stocke, J. T., Yee, H. K. C.:
    1997, \journal{ApJ}, \vol{480}, 547

    Xue, S., Zhang, Y., Chen, J.:
    2000,  \journal{ApJ}, \vol{538}, 121

    Yentis D. J., Cruddace R. G., Gursky H.:
    1992, Proceedings of the Conference on Digitised Optical Sky Surveys,
    Editors, H. T. MacGillivray, E. B. Thomson, Kluwer Academic Publishers,
    Dordrecht, Boston, MA, 67

\endreferences

}

\end{multicols}
\newpage

\begin{landscape}
\small
 \begin{minipage}{175mm}
\noindent {\bf Table 2.} New \mbox{BL~Lac}s identified by Key Field program. Table references for column nine: (1) candidate
Key Field BL~Lac; (2) Confirmed Key Field BL~Lac; (3) Confirmed BL~Lac from the
literature; (4) Schwope~et~al.~(2000); (5) Fischer~et~al.~(1998);
(6) Laurent-Muehleisen~et~al.~(1993) and Perlman~et~al.~(1996); (7)
Bauer~et~al.~(2000); (8) Cristiani~et~al.~(1995); (9) Xue~et~al.~(2000);
(10) Bade~et~al.~(1994); (11) Veron-Cetty \& Veron (2006).
\vskip0.2cm
  \begin{tabular}{@{}clcccccclcc @{}}
  \hline\noalign{\smallskip}
 (1)  & (2)           & (3)  & (4)   & (5)        & (6)       & (7)  & (8)   &(9)   & (10)             & (11) \\
Table & {\it ROSAT\/} & m$_{\rm J}$ &z & \multicolumn{2}{c}{Radio Position (J2000)} & \multicolumn{2}{c}{Radio Flux} & Ref. & PSPC & N$_{\rm H}$ \\
Index & Name          &  &   & RA & Dec & 4.8 GHz & 1.4 GHz & Flag & Count Rate & ($\times$10$^{20}$\,cm$^{-2}$) \\
      &               &  & & (h~\p0m~\p0s) &(\D\p0\p0$\arcmin$~\p0$\arcsec$)& (mJy) & (mJy) & & (cts s$^{-1}$) &\\
\noalign{\smallskip}\hline\noalign{\smallskip}
 \p01 & RX~J0040.3--2719 & 17.6 & 0.172    & 00 40 16.4 & --27 19 12 &   \p053.0$\pm$4.7 &     \p0161.0$\pm$4.2  & 3,7  & 0.31$\pm$0.03   & 1.51 \\
 \p02 & RX~J0043.4--2639 & 17.3 & 1.00\p0  & 00 43 22.6 & --26 39 07 &   \p068.0$\pm$5.9 &   \p0\p078.0$\pm$2.4  & 3,8  & 0.04$\pm$0.01   & 1.44 \\
 \p03 & RX~J0045.8--3158 & 20.1 & 0.5\p0\p0& 00 45 47.7 & --31 58 33 & \p0\p03.4$\pm$0.8 & \p0\p0\p04.1$\pm$0.6       & 1    & 0.06$\pm$0.02   & 1.98 \\
 \p04 & RX~J0049.7--4151 & 18.7 & 0.421    & 00 49 39.0 & --41 51 38 &   \p019.3$\pm$2.0 &                       & 1    & 0.08$\pm$0.02   & 3.19 \\
 \p05 & RX~J0054.8--2455 & 17.5 &          & 00 54 46.8 & --24 55 30 &   \p024.5$\pm$2.5 &   \p0\p024.3$\pm$0.9  & 1    & 0.18$\pm$0.03   & 1.56 \\
\noalign{\smallskip}
 \p06 & RX~J0059.5--3510 & 19.0 &          & 00 59 31.5 & --35 10 50 &   \p039.7$\pm$3.7 &   \p0\p080.2$\pm$2.9  & 1    & 0.12$\pm$0.02   & 2.04 \\
 \p07 &RX~J0109.9--4020$^{\ddag}$& 19.5 & 0.313    & 01 09 56.6 & --40 20 51 &   \p046.8$\pm$4.2 & \p0\p057.4$\pm$4.0$^{\dag}$   & 2,4  & 0.55$\pm$0.04   & 2.66 \\
 \p08 & RX~J0140.9--4130 & 17.8 &          & 01 40 56.7 & --41 30 12 & \p0\p01.9$\pm$0.8 &                    & 2    & 0.05$\pm$0.01   & 1.79 \\
 \p09 & RX~J0439.0--5558 & 18.7 &          & 04 39 03.0 & --55 58 38 &   \p010.7$\pm$0.6 &     $\dag$               & 2    & 0.17$\pm$0.03   & 1.65 \\
10 & RX~J0506.9--5435 & 17.2 &          & 05 06 57.8 & --54 35 03 &   \p015.3$\pm$1.3 &     $\dag$         & 2    & 0.54$\pm$0.06   & 5.49 \\
\noalign{\smallskip}
11 & RX~J0528.8--5920 & 19.2 & 1.13\p0  & 05 28 46.1 & --59 20 03 &   \p019.0$\pm$2.0 &     $\dag$         & 2    & 0.03$\pm$0.01   & 4.45 \\
12 & RX~J0543.9--5532 & 16.7 &          & 05 43 57.2 & --55 32 08 &   \p036.2$\pm$2.6 &     $\dag$         & 2,5  & 0.69$\pm$0.03   & 7.21 \\
13 & RX~J1057.8--2753 & 17.8 & 0.092    & 10 57 50.7 & --27 54 11 &   \p056.0$\pm$5.0 &   \p0\p063.8$\pm$2.0       & 3,10 & 0.15$\pm$0.02   & 5.40 \\
14 & RX~J1103.6--2329 & 16.2 & 0.186    & 11 03 37.6 & --23 29 30 &   \p058.7$\pm$5.2 &   \p0\p029.7$\pm$1.2       & 2,6  & 1.60$\pm$0.10   & 5.64 \\
15 & RX~J1357.2--0146 & 16.7 & 0.546    & 13 57 13.0 & --01 46 01 & \p0\p08.1$\pm$1.1 &   \p0\p014.9$\pm$0.6       & 1    & 0.05$\pm$0.01   & 3.48 \\
\noalign{\smallskip}
16 & RX~J1357.6+0128 & 16.7 &  0.219    & 13 57 38.7 &  +01 28 14 &   \p065.1$\pm$5.7 &   \p0\p061.3$\pm$1.9       & 2    & 0.15$\pm$0.02   & 2.36 \\
17 & RX~J2319.1--4206 & 15.0 & 0.055    & 23 19 05.8 & --42 06 49 &     207.0$\pm$17.0$^{\dag}$ &        & 3,9  & 0.12$\pm$0.03   & 1.92 \\
18 & RX~J2324.7--4041 & 15.5 &          & 23 24 44.7 & --40 40 49 &   \p033.6$\pm$3.2 &              & 3,11    & 1.50$\pm$0.10   & 1.79 \\
\noalign{\smallskip}\hline
\noalign{\smallskip}
\end{tabular}
$^{\dag}$RX~J2319.1--4206 has a ATCA 4.8~GHz flux density of 207~mJy
for the core and 237~mJy for the lob, 444~mJy in total. RX~J0109.9--4020 1.4~GHz
flux density is for ATESP J010956--402051 from Prandoni~et~al.~(2000). RX~J0439.0--5558 has a
843~MHz flux density of 21.2$\pm$1.3~mJy for SUMSS J043901--555840 from Mauch~et~al.~(2003), 
and includes RX~J0506.9--5435 with 17.1$\pm$1.1~mJy, RX~J0528.8--5920 with 18.8$\pm$1.0~mJy,
RX~J0543.9--5532 with 41$\pm$2~mJy.

$^{\ddag}$For more details see Section~4.1.
\end{minipage}
\end{landscape}

\vfill\eject

{\ }



\naslov{RADIO DETEKCIJA 18 {\rm RASS \mbox{BL~LAC}} OBJEKATA}


\authors{M. W. B. Anderson$^{\bf 1,2}$ and M. D. Filipovi\'c$^{\bf 2}$ }

\vskip3mm

\address{$^1$Sydney Observatory, PO Box K346, Haymarket, Sydney, NSW 1238, Australia}

\address{$^2$University of Western Sydney, Locked Bag 1797\break Penrith South, DC, NSW 1797, Australia}

\vskip3mm


\centerline{\rrm UDK \udc}

\vskip1mm

\centerline{\rit Originalni nauqni rad}

\vskip.7cm

\begin{multicols}{2}

{


\rrm

U ovoj studiji predstav{lj}amo radio detekciju 18 novih {\rm \mbox{BL~Lac}} objekata sa naxeg pregleda neba od preko 575 kvadratnih stepeni. Ovih 18 objekata su locirani u okolini od $20\arcsec$ od {\rm \mbox{X-ray}} pozicije. Za 11 od 18 objekata imamo izmereni crveni pomak. Svih 18 kandidata su radio emiteri iznad {\rm $\sim$1~mJy} i podpadaju u poznatu grupaciju od ``dve boje'', {\rm $\alpha_{\rm ro}$ vs $\alpha _{\rm ox}$}, diagram sa evidentnom tranzicionom populacijom 3. Detektovali smo i dva veoma neobiqna izvora: kandidat za radio tihu {\rm BL~Lac, RX~J0140.9--4130,} i ekstremni HBL, {\rm \mbox{RX~J0109.9--4020}}, sa {\rm $Log\left( \nu _{\rm peak} \right)\approx19.2$.}  {\rm \mbox{BL~Lac}~Log(N)--Log(S)} relacija je konsistentna sa ostalim uzorcima i indicira da \textit{ROSAT} {\rm All Sky Survey (RASS)} ima i do ($2000{\pm}400$) {\rm \mbox{BL~Lac}} objekata.

}

\end{multicols}

\end{document}